\documentclass[submission,copyright,creativecommons]{eptcs}
\providecommand{\event}{GCM 2025}

\usepackage{iftex}

\ifpdf
  \usepackage{underscore}         
  \usepackage[T1]{fontenc}        
\else
  \usepackage{breakurl}           
\fi

\usepackage[ruled,vlined]{algorithm2e}
\usepackage{algorithmic}
\usepackage{amsfonts}
\usepackage{amsmath}
\usepackage{amsthm}
\usepackage{multicol}
\usepackage{wrapfig}
\usepackage{setspace}
\usepackage{graphicx}
\usepackage{imakeidx}
\usepackage{scalerel}
\usepackage{tikz}
\usepackage{txfonts}

\tikzset{ 
ev/.pic = { \fill[black] (0.0, 0.0) circle (1.0mm) ; } ,
iv/.pic = { \filldraw[black, fill=white, line width=0.5mm] circle (.8mm) ; } ,
de/.style = { black, line width=0.5mm, -> } ,
ue/.style = { black, line width=0.5mm } ,
he/.style = { black, line width=1.0mm } ,
hl/.style = { fill=white, draw=black, line width=0.5mm, rounded corners=0.1mm } ,
ht/.style = { fill=graph_term, draw=black, line width=0.25mm, rounded corners=0.1mm } ,
el/.style = { text width = 1.5mm, text height = 1.5mm, font = \scriptsize, align=center} ,
pn/.style = { text width = 10.0mm, align=center } ,
hn/.style = { text width = 3.0mm, align=center } ,
tn/.pic = { \draw[fill=white, draw=black, line width=0.5mm] circle (30.0mm) ; } ,
cn/.pic = { \draw[fill=white, draw=black, line width=0.5mm] circle (5.0mm) ; } ,
te/.style = { fill=white } ,
tb/.style = { draw=black, line width=0.25mm } ,
sl/.style = { sloped, scale=2, pos=0.5, allow upside down }
}

\definecolor{graph_term}{RGB}{223, 223, 223}

\newcommand\psb[3]{#1^{#2}\mathord{\_}#3}
\newcommand\psa{\mathrel{\land}}
\newcommand\pse[1]{\ #1\ }

\title{Parsing Hypergraphs using\\ Context-Free Positional Grammars}
\author{Gennaro Costagliola\thanks{\href{https://orcid.org/0000-0003-3816-7765}{G. Costagliola ORCID: 0000-0003-3816-7765}}
\institute{University of Salerno}
\email{gencos@unisa.it}
\and
Federico Vastarini\thanks{\href{https://orcid.org/0000-0002-8002-5704}{F. Vastarini ORCID: 0000-0002-8002-5704} (corresponding author)}
\institute{University of Salerno}
\email{fvastarini@unisa.it}
}
\def\titlerunning{Parsing Hypergraphs using Context-Free Positional Grammars}
\def\authorrunning{Gennaro Costagliola, Federico Vastarini}
\begin{document}
\maketitle
\let\thefootnote\relax\footnotetext{Authors are listed in alphabetical order.}

\begin{abstract}
We present a novel work-in-progress approach to the parsing of hypergraphs generated by context-free hyperedge replacement grammars. This method is based on a new LR parsing technique for positional grammars, which is also under active development. Central to our approach is a reduction from hyperedge replacement to positional grammars with additional structural constraints, enabling the use of permutation-based operations to determine the correct ordering of hyperedges on the right-hand side of productions. Preliminary results also reveal a distinction between ambiguity in graph generation and ambiguity in graph recognition. While the exact class of hyperedge replacement languages parsable under this method remains under investigation, the approach provides a promising foundation for future generalisations to more expressive grammar formalisms. Graph parsing remains a broadly relevant problem across numerous domains, and our contribution aims to advance both the theoretical and practical understanding of this challenge.
\end{abstract}

\paragraph{Introduction:}
\label{par:introduction}
Parsing hypergraphs in context-free \cite{chomsky-1959-ocf,courcelle-1987-aad} hyperedge replacement languages (\textit{HRLs}) \cite{drewes-1997-hrg,engelfriet-1997-cfg} is a well-established yet computationally challenging problem. While this problem is thoroughly explored in \cite{drewes-2019-fac}, where Predictive Shift-Reduce (\textit{PSR}) parsers were introduced as a graph-analogue of SLR(1) string parsing, this work-in-progress proposes a novel approach to building an efficient parser for a class of such languages that is currently under investigation. The key idea is to exploit the ordering of hyperedges on the right-hand side of productions, as introduced in \cite{vastarini-2024-rgg} for uniform graph generation, and adapt it to the parsing domain. The approach is divided into two parts. It begins with a translation of hyperedge replacement grammars (\textit{HRGs}) into positional grammars (\textit{PGs}) \cite{costagliola-1997-apm,costagliola-2020-rtc}. Then, a method is proposed in which the resulting grammar is restructured and reordered to potentially eliminate parsing conflicts and enable efficient parsing. The outcome of the method is an item-based parser \cite{earley-1970-aep} employing dot operators. Positional strings representing hypergraphs are parsed using adapted \textit{LR} techniques \cite{knuth-1965-ott}. Our proposal is based on a bottom-up parser and does not yet consider top-down parsers, as discussed in \cite{drewes-2020-gpa}, and another related approach, which uses node ordering in a different class of hypergraphs, is proposed in \cite{bjorklund-2021-upf}.
  
\paragraph{Hyperedge Replacement Grammars:}
\label{par:hyperedgeReplacementGrammars}
We begin by concisely providing the necessary foundations for our formalism of context-free \textit{HRGs}, ensuring that key properties are preserved throughout the translation process. A thorough description of \textit{HRLs} can be found in \cite{vastarini-2024-rg1,drewes-1997-hrg,engelfriet-1997-cfg}. Labelled hypergraphs are a generalisation of standard graphs, where each edge, referred to as a hyperedge, carries a label. Formally, let $\textit{type} \colon C \to \mathbb{N}_0$ be a typing function for a fixed set of labels $C$, then a \textit{hypergraph} over $C$ is a tuple $H=(V_H,E_H,\textit{att}_H,\textit{lab}_H,\textit{ext}_H)$ where $V_H$ is a finite set of \textit{vertices}, $E_H$ is a finite set of \textit{hyperedges}, $\textit{att}_H\colon E_H \to V^*_H$ is a mapping assigning a sequence of \textit{attachment nodes} to each $e \in E_H$, $\textit{lab}_H\colon E_H \to C$ is a function that maps each hyperedge to a \textit{label} such that $\textit{type}(\textit{lab}_{H}(e)) = |\textit{att}_{H}(e)|$, $\textit{ext}_H \in V^*_H$ is a sequence of pairwise distinct \textit{external nodes}. Note that external nodes are only used during the derivation of a hypergraph. We write $\textit{type}(H)$ for $|\textit{ext}_{H}|$. The length of the sequence of \textit{attachments} $|\textit{att}_{H}(e)|$ is the \textit{type} of $e$. In the context of this work, we are primarily interested in hypergraphs without isolated nodes or hyperedges. That is, each hyperedge must be of at least \textit{type}-$1$ and each node must have at least one adjacent edge. External nodes are depicted with filled circles. Elementary operations for replacing hyperedges with hypergraphs of matching \textit{type} are defined via \textit{productions}. Let $N \subseteq C$ be a subset of non-terminal labels. Then $p = (A,R)$ is a \textit{production} over $N$, where $\textit{lhs}(p) = A \in N$ is the label of the replaced hyperedge and $\textit{rhs}(p) = R$ is a hypergraph with $\textit{type}(R) = \textit{type}(A)$. We use lowercase letters to denote terminal hyperedges, and uppercase letters to denote non-terminal ones.

Given an ordered set $\{\alpha_1, \ldots, \alpha_n\}$ where $\alpha_i < \alpha_j$ if $i < j \in \mathbb{N}$ we define a \textit{hyperedge replacement grammar}, or \textit{HRG} as a tuple $G = (N,\Sigma,P,S,(\textit{mark}_{p})_{p \in P})$ where $N \subseteq C$ is a finite set of non-terminal labels, $\Sigma \subseteq C$ is a finite set of terminal labels with $N \cap \Sigma = \emptyset$, $P$ is a finite set of productions, $S \in N$ is the starting symbol, $(\textit{mark}_{p})_{p \in P}$ is a family of functions $\textit{mark}_p: E_R \to \{\alpha_1, \ldots, \alpha_n\}$ assigning a mark to each hyperedge in the right-hand side of a production $p$. For each pair $e_i, e_j \in E_R$ with $i \neq j$, $\textit{mark}(e_i) \neq \textit{mark}(e_j)$. This ordering is treated as a property of the grammar, such that grammars with different orderings are considered distinct. It is always possible to assign an arbitrary order to the hyperedges on the \textit{rhs} of productions of an \textit{HRG}, and our method aims to find an ordering, if it exists, that allows the generation of a \textit{well-formed} grammar.

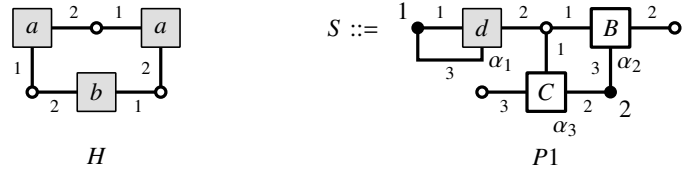
\begin{figure}[htpb]
\centering
\resizebox{9.0cm}{!}{%
\begin{tikzpicture}
 \begin{scope} [local bounding box = c1, shift = {(0.0, 0.0)}]
  \draw [ue] (0.0, -0.3) -- (0.0, -1.0) node[el, midway, left] {1} ;
  \draw [ue] (0.3, 0.0) -- (1.0, 0.0) node[el, midway, above] {2} ;
  \draw [ue] (1.7, 0.0) -- (1.0, 0.0) node[el, midway, above] {1} ;
  \draw [ue] (2.0, -0.3) -- (2.0, -1.0) node[el, midway, left] {2} ;
  \draw [ue] (1.3, -1.0) -- (2.0, -1.0) node[el, midway, below] {1} ;
  \draw [ue] (0.7, -1.0) -- (0.0, -1.0) node[el, midway, below] {2} ;
  \draw [ht] (-0.3, 0.3) rectangle (0.3, -0.3) node[midway] {$a$} ;
  \draw [ht] (1.7, 0.3) rectangle (2.3, -0.3) node[midway] {$a$} ;
  \draw [ht] (0.7, -0.7) rectangle (1.3, -1.3) node[midway] {$b$} ;
  \path (1.0, 0.0) pic{iv} ;
  \path (2.0, -1.0) pic{iv} ;
  \path (0.0, -1.0) pic{iv} ;
  \node [pn] at (1.0, -2.0) {$H$} ;
 \end{scope}
 
 \node[text width = 1.0cm, align=center] at (5.0, 0.0) {$S::=$} ;
 
 \begin{scope} [local bounding box = c1, shift = {(6.0, 0.0)}]
  \draw [ue] (0.7, 0.0) -- (0.0, 0.0) node[el, midway, above] {1} ;
  \draw [ue] (1.3, 0.0) -- (2.0, 0.0) node[el, midway, above] {2} ;
  \draw [ue] (1.0, -0.3) -- (1.0, -0.5) -- (0.0, -0.5) node[el, midway, below] {3} -- (0.0, 0.0) ;
  \draw [ue] (2.0, 0.0) -- (2.7, 0.0) node[el, midway, above] {1} ;
  \draw [ue] (3.3, 0.0) -- (4.0, 0.0) node[el, midway, above] {2} ;
  \draw [ue] (3.0, -0.3) -- (3.0, -1.0) node[el, midway, left] {3} ;
  \draw [ue] (2.0, -0.7) -- (2.0, 0.0) node[el, midway, right] {1} ;
  \draw [ue] (2.3, -1.0) -- (3.0, -1.0) node[el, midway, below] {2} ;
  \draw [ue] (1.7, -1.0) -- (1.0, -1.0) node[el, midway, below] {3} ;
  \draw [ht] (0.7, 0.3) rectangle (1.3, -0.3) node[midway] {$d$} node[below] {$\alpha_1$} ;
  \draw [hl] (2.7, 0.3) rectangle (3.3, -0.3) node[midway] {$B$} node[below] {$\alpha_2$} ;
  \draw [hl] (1.7, -0.7) rectangle (2.3, -1.3) node[midway] {$C$} node[below] {$\alpha_3$} ;
  \path (0.0, 0.0) pic{ev} node[above left] {1} ;
  \path (2.0, 0.0) pic{iv} ;
  \path (4.0, 0.0) pic{iv} ;
  \path (1.0, -1.0) pic{iv} ;
  \path (3.0, -1.0) pic{ev} node[below right] {2} ;
  \node [pn] at (2.0, -2.0) {$P1$} ;
 \end{scope}
\end{tikzpicture}%
}
\caption[Hyperedge and Production]{Example of a hypergraph and an \textit{HRG} production}
\label{fig:hypergraphs}
\end{figure}

The hypergraph $H$ in Figure \ref{fig:hypergraphs} is an example of a cycle graph with three terminal type-$2$ hyperedges labelled $a$, $a$ and $b$. This hypergraph, without markings and external nodes, is an example of input for the parser. Note, for example, that cycle graphs do not have an automatically recognisable starting point. Production $P1$, in the same figure, describes the replacement of a hyperedge labelled with $S$ on its \textit{lhs} with a hypergraph of the same type on its \textit{rhs}. The \textit{type}-$2$ hypergraph on the \textit{rhs} has $5$ nodes, of which $2$, depicted with full circles, are external. The hyperedge labelled with $d$ is a terminal hyperedge, while the others, labelled with $B$ and $C$, are non-terminal. The marks $\alpha_1$, $\alpha_2$ and $\alpha_3$ represent a possible ordering in which the hyperedges are considered during parsing. There are $n!$ possible orderings, where $n$ is the number of hyperedges.

\paragraph{From \textit{HRGs} to Positional Grammars:}
\label{par:fromHRGsToPositionalGrammars}
We concisely introduce context-free \textit{PGs} and describe how to represent an \textit{HRG} within this framework. A \textit{positional string} $s = c_1 e_1 \ldots c_n e_n$ is an ordered sequence of \textit{elements}, each preceded by a \textit{conjunction of connectors}. We refer to the element following the conjunction as its \textit{target}. Each element is characterized by a \textit{symbol} and a sequence of \textit{interfaces}. Each connector in a conjunction describes a binary relation between an interface of a \textit{source} element and an interface of the target element. Such relations may also involve two different interfaces of the same element. A connector is denoted as $\psb{k}{z}{l}$, where $k$ is the interface of the source, $l$ is the interface of the target. To ensure compatibility with the existing notation, we define $z$ as the distance of the source with respect to the target, decremented by one. That is, $z=-1$ denotes that the source coincides with the target, while $z=0$ refers to the element immediately preceding it, and so on. By convention, $z=0$ is assumed by default and may be omitted. In order to represent nodes connected to a single hyperedge we also use a \textit{unary} relation that involves only an interface of a single element. For example, the second attachment point of the $B$-labelled hyperedge in the \textit{rhs} of $P1$ in Figure \ref{fig:hypergraphs}. In order to comply with the form of the current input of the parser, unary relations are not explicitly shown in conjunctions, but they are inferred from the description of the elements used in the \textit{rhs} of the productions of the grammar. For example, the production in Figure \ref{fig:hypergraphs} corresponds to the following positional production:

$S \Rightarrow \psb{1}{-1}{3} \pse{d} \psb{2}{}{1} \pse{C} \psb{2}{1}{1} \psa \psb{3}{}{2} \pse{B}(S1 = d1, S2 = C2)$

To enable the representation of \textit{HRGs} using \textit{PGs}, we consider invertible relations that satisfy the following properties:
\begin{itemize}
\itemsep0em
\item[$\triangleright$] \textit{Uniqueness:} Each interface must participate in exactly one type of relation.
\item[$\triangleright$] \textit{Symmetry:} Binary relations are symmetric: if a relation exists from interface $1$ to interface $2$, then the same relation also exists from $2$ to $1$.
\item[$\triangleright$] \textit{Transitivity:} Binary relations are transitive: if a relation exists from interface $1$ to interface $2$, and the same relation exists from $2$ to another interface $3$, then it must also exist from $1$ to $3$.
\end{itemize}

To describe the structure of a hypergraph, we use the binary relation \textit{``shares"} to represent either two distinct hyperedges that are adjacent to the same node, or a single hyperedge with two distinct attachment points connected to the same node. When two or more hyperedges are adjacent to the same node, this forms a relation that is both symmetric and transitive. The same holds when two or more attachment points of a single hyperedge are connected to the same node. The restriction of uniqueness ensures that an interface may be involved in binary or unary relations, but not both at the same time. For example, the hypergraphs in Figure \ref{fig:hypergraphs} correspond to the following positional strings: $\pse{a} \psb{2}{}{1} \pse{a} \psb{1}{1}{2} \psa \psb{2}{}{1} \pse{b}$.

\paragraph{Parser:}
\label{par:parser}
The parser follows the principles of the \textit{pLR} parsing methodology, as described in \cite{costagliola-1997-apm, costagliola-2020-rtc}. This is an extension of \textit{LR} parsing in which the construction of \textit{LR} items considers positional strings instead of traditional symbolic strings. A dotted symbol in an item indicates which symbol to expect, and a dotted conjunction of connectors indicates where to retrieve the next symbol. This parsing methodology applies to positional grammars that are \textit{well-formed}, i.e., they respect the restriction defined in \cite{costagliola-1997-apm}, which requires that, for each production, each entering attachment point of the \textit{lhs} non-terminal is tied to an attachment point of the leftmost symbol of the \textit{rhs}. A similar requirement exists in the \textit{LR}-based parsing methodology proposed in \cite{drewes-2019-fac}, where the grammar designer must carefully devise the production order to ensure the parser functions correctly. Compared to this approach, among other differences, the \textit{pLR} parsing methodology enables the construction of an \textit{LR}-based parser for structured flowcharts, as shown in \cite{costagliola-1994-tep}, by guaranteeing that the construction of the item sets is always a finite process.

The approach we propose removes the constraint of requiring well-formed grammars as input by introducing a grammar transformation procedure capable of correctly permuting the order of symbols within productions. This procedure, the limits of which are currently under investigation, can be applied either as an initial step of the methodology or during the creation of item sets, thus granting language designers greater flexibility in defining grammars. It can be seen as a complementary path to the \textit{PSR} method in \cite{drewes-2019-fac} toward deterministic \textit{LR-style} parsing of hyperedge replacement languages. While \textit{PSR} preserves the \textit{HRG} formalism and enforces determinism through global conditions such as viable prefixes and the free edge choice property, our approach translates \textit{HRGs} into positional form and enforces determinism through suitable permutations of productions. Together, these perspectives highlight the growing convergence between string-based \textit{LR} techniques and their graph-based generalisations.

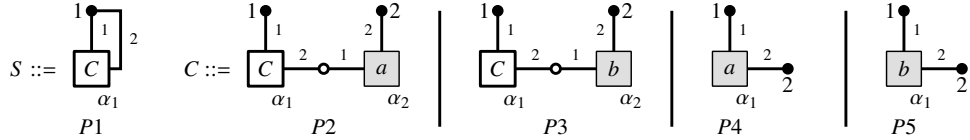
\begin{figure}[htpb]
\centering
\resizebox{13.0cm}{!}{%
\begin{tikzpicture}
 \node[text width = 1.0cm, align=center] at (0.0, 0.0) {$S::=$} ;
 
 \begin{scope} [local bounding box = c1, shift = {(1.0, 0.0)}]
  \draw [ue] (0.0, 0.3) -- (0.0, 1.0) node[el, midway, right] {1} ;
  \draw [ue] (0.3, 0.0) -- (0.5, 0.0) -- (0.5, 1.0) node[el, midway, right] {2} -- (0.0, 1.0) ;
  \draw [hl] (-0.3, 0.3) rectangle (0.3, -0.3) node[midway] {$C$} node[below] {$\alpha_1$} ;
  \path (0.0, 1.0) pic{ev} node[left] {1} ;
  \node [pn] at (0.0, -1.0) {$P1$} ;
 \end{scope}
 
 \node[text width = 1.0cm, align=center] at (3.0, 0.0) {$C::=$} ;
 
 \begin{scope} [local bounding box = c1, shift = {(4.0, 0.0)}]
  \draw [ue] (0.0, 0.3) -- (0.0, 1.0) node[el, midway, right] {1} ;
  \draw [ue] (0.3, 0.0) -- (1.0, 0.0) node[el, midway, above] {2} ;
  \draw [ue] (1.7, 0.0) -- (1.0, 0.0) node[el, midway, above] {1} ;
  \draw [ue] (2.0, 0.3) -- (2.0, 1.0) node[el, midway, left] {2} ;
  \draw [hl] (-0.3, 0.3) rectangle (0.3, -0.3) node[midway] {$C$} node[below] {$\alpha_1$} ;
  \draw [ht] (1.7, 0.3) rectangle (2.3, -0.3) node[midway] {$a$} node[below] {$\alpha_2$} ;
  \path (0.0, 1.0) pic{ev} node[left] {1} ;
  \path (2.0, 1.0) pic{ev} node[right] {2} ;
  \path (1.0, 0.0) pic{iv} ;
  \node [pn] at (1.0, -1.0) {$P2$} ;
 \end{scope}
 
 \draw[black, line width=0.5mm] (7.0, 1.0) -- (7.0, -1.0) ;
 
 \begin{scope} [local bounding box = c1, shift = {(8.0, 0.0)}]
  \draw [ue] (0.0, 0.3) -- (0.0, 1.0) node[el, midway, right] {1} ;
  \draw [ue] (0.3, 0.0) -- (1.0, 0.0) node[el, midway, above] {2} ;
  \draw [ue] (1.7, 0.0) -- (1.0, 0.0) node[el, midway, above] {1} ;
  \draw [ue] (2.0, 0.3) -- (2.0, 1.0) node[el, midway, left] {2} ;
  \draw [hl] (-0.3, 0.3) rectangle (0.3, -0.3) node[midway] {$C$} node[below] {$\alpha_1$} ;
  \draw [ht] (1.7, 0.3) rectangle (2.3, -0.3) node[midway] {$b$} node[below] {$\alpha_2$} ;
  \path (0.0, 1.0) pic{ev} node[left] {1} ;
  \path (2.0, 1.0) pic{ev} node[right] {2} ;
  \path (1.0, 0.0) pic{iv} ;
  \node [pn] at (1.0, -1.0) {$P3$} ;
 \end{scope}
 
 \draw[black, line width=0.5mm] (11.0, 1.0) -- (11.0, -1.0) ;
 
 \begin{scope} [local bounding box = c1, shift = {(12.0, 0.0)}]
  \draw [ue] (0.0, 0.3) -- (0.0, 1.0) node[el, midway, right] {1} ;
  \draw [ue] (0.3, 0.0) -- (1.0, 0.0) node[el, midway, above] {2} ;
  \draw [ht] (-0.3, 0.3) rectangle (0.3, -0.3) node[midway] {$a$} node[below] {$\alpha_1$} ;
  \path (0.0, 1.0) pic{ev} node[left] {1} ;
  \path (1.0, 0.0) pic{ev} node[below] {2} ;
  \node [pn] at (0.0, -1.0) {$P4$} ;
 \end{scope}
 
 \draw[black, line width=0.5mm] (14.0, 1.0) -- (14.0, -1.0) ;
 
 \begin{scope} [local bounding box = c1, shift = {(15.0, 0.0)}]
  \draw [ue] (0.0, 0.3) -- (0.0, 1.0) node[el, midway, right] {1} ;
  \draw [ue] (0.3, 0.0) -- (1.0, 0.0) node[el, midway, above] {2} ;
  \draw [ht] (-0.3, 0.3) rectangle (0.3, -0.3) node[midway] {$b$} node[below] {$\alpha_1$} ;
  \path (0.0, 1.0) pic{ev} node[left] {1} ;
  \path (1.0, 0.0) pic{ev} node[below] {2} ;
  \node [pn] at (0.0, -1.0) {$P5$} ;
 \end{scope}
\end{tikzpicture}%
}
\caption[Hypergraphs]{Cycle graph \textit{HRG}}
\label{fig:cycleGraphGrammar}
\end{figure}

To support the description of our parsing method, we introduce an example based on a cycle-graph grammar.
The following \textit{PG} corresponds to the cycle-graph \textit{HRG} depicted in Figure \ref{fig:cycleGraphGrammar}, where the order of elements reflects the order of the hyperedges.

$S \longrightarrow \psb{2}{-1}{1} \pse{C} (S1 = C1) \ \ \ \ C \longrightarrow \pse{C} \psb{2}{}{1} \pse{a} \ (C1 = C1, C2 = a2)$

$C \longrightarrow \pse{C} \psb{2}{}{1} \pse{b} \ (C1 = C1, C2 = b2) \ \ \ \ C \longrightarrow \pse{a} \ (C1 = a1, C2 = a2) \ \ \ \ C \longrightarrow \pse{b} \ (C1 = b1, C2 = b2)$

Given a \textit{PG}, an equivalent well-formed grammar can be obtained by applying a series of permutations to its productions:
\begin{itemize}
\itemsep0em
\item[$\triangleright$] \textit{Simple permutation:} A rearrangement of the sequence of elements on the \textit{rhs} of a production. Since changing the order of elements on the \textit{rhs} results in a representation of an isomorphic hypergraph with a different ordering, it affects only how the string is traversed during parsing.
\item[$\triangleright$] \textit{Duplicated permutation:} A rearrangement of the elements, as in the previous case, but with a copy of the original production also retained in the grammar. This technique ensures that the production can be processed in both orders, at the cost of an increased grammar size.
\end{itemize}

The above \textit{PG} is an example of a well-formed grammar that yields an efficient, conflict-free \textit{pLR} parser for cycle-graphs, regardless of the element from which parsing starts. In our approach, the ordering of hyperedges in suitable permutations of productions plays a role similar to that of the \textit{Follow} and $\textit{Follow}^*$ sets in \textit{PSR} parsing, guiding the parser toward deterministic choices. Moreover, just as \textit{PSR} relies on the \textit{free edge choice property} to guarantee that any fitting edge can be shifted without failure, our use of permutations ensures that whichever hyperedge the parser encounters first, the derivation can proceed without backtracking.

The cycle graph \textit{PG} is no longer well-formed if, for example, $P1$ is permuted into:

$C \longrightarrow \pse{a} \psb{1}{}{2} \pse{C} \ \ (C1 = C1, C2 = a2)$

In this case, the entering attaching point $1$ of the \textit{lhs} non-terminal of the production is not tied to any attaching point of $a$, the leftmost symbol of the \textit{rhs} of the production. A related notion in the \textit{PSR} framework is that of viable prefixes. In our setting, well-formedness plays a similar role, ensuring that prefixes never leave dangling or unanchored attachment points.

\paragraph{Conclusion:}
\label{par:conclusion}
The goals of this work currently focus on the following objectives:

\textit{Efficient Algorithm:} Since the behaviour of the parser is directly influenced by the ordering of the elements, an efficient algorithm that minimises the number of duplicated permutations is essential for efficient parsing. Ideally, the correct order should follow from an analysis of the grammar’s structure, identifying how external nodes and attachment points are introduced and propagated across productions. Such an analysis would allow the parser to determine systematically which hyperedges should appear first on the right-hand side, ensuring well-formedness without resorting to arbitrary or exhaustive reordering.

\textit{Class of \textit{HRLs}:} Another possible extension is to study the class of \textit{HRLs} to which this parsing method applies, through the use of \textit{PGs}.

\bibliographystyle{eptcs}
\bibliography{common}

\end{document}